\documentclass[aps,prd,preprint,tightenlines,floatfix]{revtex4}
\usepackage{graphicx}
\usepackage{wrapfig}
\usepackage[all]{xy}
\newcommand{\nc}{\newcommand}

\nc{\beq}{\begin{equation}}
\nc{\eeq}{\end{equation}}
\nc{\bea}{\begin{eqnarray}}
\nc{\eea}{\end{eqnarray}}
\nc{\n}{\nonumber \\}

\begin{document}  

\date{Jan 28, 2011} 
\title{DAMA and the self similar infall halo model}
\author{Aravind Natarajan}
\email{anat@andrew.cmu.edu}
\affiliation{McWilliams Center for Cosmology, Carnegie Mellon University, Department of Physics, 5000 Forbes Ave., Pittsburgh PA 15213, USA}

\begin{abstract} 
The annual modulation in the rate of WIMP recoils observed by the DAMA collaboration at high significance is often analyzed in the context of an isothermal Maxwell-Boltzmann velocity distribution. While this is the simplest model, there is a need to consider other well motivated theories of halo formation. In this paper, we study a different halo model, that of self similar infall which is characterized by the presence of a number of cold streams and caustics, not seen in simulations. It is shown that the self similar infall model is consistent with the DAMA result both in amplitude and in phase, for WIMP masses exceeding $\approx$ 250 GeV at the 99.7\% confidence level. Adding a small thermal component makes the parameter space near $m_\chi$ = 12 GeV consistent with the self similar model. The minimum $\chi^2$ per degree of freedom is found to be 0.92(1.03) with(without) channeling taken into account, indicating an acceptable fit. For WIMP masses much greater than the mass of the target nucleus, the recoil rate depends only on the ratio $\sigma_{\rm p}/m_\chi$ which is found to be $\approx$ 0.06 femtobarn/TeV. However as in the case of the isothermal halo, the allowed parameter space is inconsistent with the null result obtained by the CDMS and Xenon experiments for spin-independent elastic scattering. Future experiments with directional sensitivity and mass bounds from accelerator experiments will help to distinguish between different halo models and/or constrain the contribution from cold flows.
\end{abstract}

\maketitle

\section{Introduction}

It was shown by Drukier, Freese, and Spergel \cite{drukier}, and by Freese, Frieman, and Gould \cite{freese} that the motion of the earth about the sun introduces an annual modulation in the flux of dark matter particles reaching the earth. The detection of such an annual modulation has been claimed by the DAMA/NaI and DAMA/LIBRA experiments \cite{dama1, dama2} conducted at the Gran Sasso National Laboratory using highly pure NaI(Tl) detectors. The DAMA experiment has reported its results for a cumulative time period of 13 annual cycles and a total exposure of 1.17 ton-year, claiming a detection of the annual modulation signature at $> 8 \sigma$ \cite{dama1}. The DAMA claim is strengthened by the fact that only single hit events (expected to be triggered by particles with a weak cross section) are annually modulated, the multiple hit events show no statistically significant modulation. We refer the reader to \cite{dama1, dama2} for details regarding the experimental setup and backgrounds.

The annual modulation seen by the DAMA experiment is commonly analyzed in the context of an isothermal Maxwell-Boltzmann velocity distribution implying a WIMP mass $m_\chi \approx 12$ GeV or $m_\chi \approx 78$ GeV for the simple case of elastic spin-independent scattering. The derived values of mass and cross section are inconsistent with the null result obtained by other dark matter direct detection experiments such as CDMS \cite{cdms} and Xenon \cite{xenon}, for the simple case of spin-independent elastic scattering. The low mass region may also be challenged by observations of the CMB \cite{cirelli, galli, arvi_reion}, or by future accelerator experiments.

The purpose of this paper is to compare the DAMA results to a non-standard halo model, namely self similar infall. The self similar infall halo is characterized by a number of discrete cold flows and caustics, not seen in numerical simulations. The presence of a cold flow is significant since dark matter detection experiments such as DAMA are sensitive to the local phase space distribution. The annual modulation effect predicted by the self similar infall halo model was studied by Copi and Krauss \cite{copi}, Green \cite{green}, Gelmini and Gondolo \cite{gelmini}, Vergados \cite{vergados}, and Ling, Sikivie, and Wick \cite{ling}. In these papers, it was shown that the self similar model predicts qualitatively different results than those predicted by the Maxwellian halo. However this does not mean that the maximum recoil rate observed by DAMA on May 25 $\pm$ 8 days in the $2-6$ keV$_{\rm ee}$ range \cite{dama2} is inconsistent with the self similar infall model. We show here that for WIMP masses exceeding 250 GeV, the self similar model is in agreement with the DAMA observation at the 99.7\% level.  Nevertheless as in the case of the isothermal halo, the allowed parameter space is in contradiction with the exclusion limits obtained by other experiments for spin-independent elastic scattering.

In section II, we derive the recoil rate observed by the DAMA experiment in terms of the mass, cross section and velocity distribution. We then briefly discuss the self similar infall halo model in Section III. The model is characterized by a  series of cold flows, one of which is dominant due to the presence of a nearby dark matter caustic. The fractional density contributed by the dominant flow is fixed by requiring that the recoil rate be a maximum on the observed date of May 25 $\pm$ 8 days.   We discuss the effect of the dominant flow on the annual modulation for Na and I nuclei. We then present our results in Section IV. We present the spectrum of expected recoil events for different times of the year, for 4 different energy bins. A $\chi^2$ analysis is performed to determine the best fit values of mass and cross section. We show that the $\chi^2$ per degree of freedom is close to unity, indicating an acceptable fit. We compare the results with that of a Maxwellian halo, and also with the CDMS and Xenon bounds. We then check that introducing a small thermal component does not lead to qualitatively different predictions. Finally, we verify that the time averaged recoil rate predicted by the self-similar model is consistent with the average plus background reported by the DAMA collaboration. In Section V, we summarize our results and discuss various ways in which the model may be tested by future experiments.

\section{Recoil rate.}

\begin{figure}[!h]
\begin{center}
\scalebox{0.6}{\includegraphics{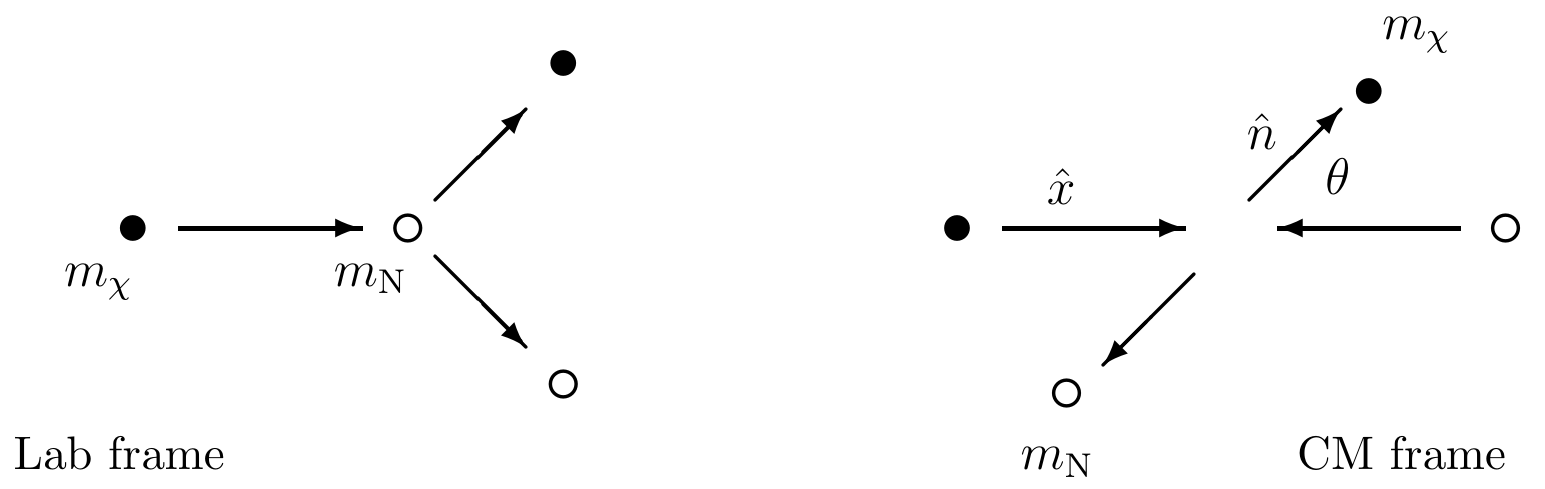}}
\end{center}
\caption{  WIMP-nucleus scattering.   \label{fig1} }
\end{figure}
Consider an elastic collision between a dark matter particle with mass $m_\chi$ and a target nucleus of mass $m_{\rm N}$. The dark matter particle has a  velocity $\vec v = v \, \hat x$ relative to the target nucleus. The velocity of the center of momentum is given by $\vec v_{\rm cm} = \frac{ m_\chi v }{m_\chi + m_{\rm N}} \hat x$. The velocity of the recoiling nucleus in the CM frame is $\vec v'_{\rm N, CM} = - \frac{ m_\chi v }{m_\chi + m_{\rm N}} \hat n$. Therefore, in the lab frame (where the detector is at rest), the recoil velocity is $ \vec v'_{\rm N, lab} = \vec v'_{\rm N, CM} + \vec v_{\rm CM} = \frac{ m_\chi v }{m_\chi + m_{\rm N}} (\hat x - \hat n)$. The kinetic energy of the recoiling nucleus in the lab frame is:
\beq
Q = \frac{m_{\rm N} v'^{ 2}_{\rm N, lab}}{2} =  \frac{m^2_{\rm R} v^2 }{m_{\rm N}} \left(1 - \cos\theta \right ),
\eeq
where $\theta$ is the scattering angle in the CM frame, and $m_{\rm R} = m_\chi m_{\rm N} / (m_\chi + m_{\rm N})$ is the WIMP-nucleus reduced mass.  The maximum possible recoil energy when the WIMP has a speed $v$ relative to the detector is obtained when $\theta = \pi$:
\beq
Q_{\rm max} = \frac{2 m^2_{\rm R} v^2}{m_{\rm N}},
\eeq
and therefore, the minimum velocity the WIMP must have in order to effect a recoil at energy $Q$ is found to be
\beq
v_{\rm min} = \sqrt{ \frac{Q m_{\rm N}}{2 m^2_{\rm R}} }.
\eeq
For example, a 100 GeV WIMP moving at a speed $10^{-3}$ c and colliding with a $\approx$ 120 GeV Iodine nucleus deposits energy $\approx$ 25 keV $(1-\cos\theta)$.

The number of recoil events seen by the detector per unit time per unit detector mass and per unit recoil energy is
\beq
\frac{dR}{dQ} = \frac{1}{m_{\rm N}} \, \frac{\rho_\chi}{m_\chi} \, \int_{v_{\rm min}(Q)}^\infty  \frac{d\sigma}{dQ} \; v f(v) dv.
\eeq
$\sigma$ is the WIMP-nucleus scattering cross section, $\rho_\chi$ is the dark matter density at the earth's location, and $f(v)$ is the speed distribution of WIMPs relative to the detector. The differential cross section is commonly expressed as:
\bea
\frac{d\sigma}{dQ} &=& \frac{\sigma_0}{Q_{\rm max}} F^2(Q) \n
&=& \frac{\sigma_0 m_{\rm N}}{2 m^2_{\rm R} v^2} F^2(Q).
\eea
$F(Q)$ is called the form factor and contains the momentum dependence of the cross section. We assume $F(Q)$ may be described by the form \cite{form_factor1,form_factor2,form_factor3}
\beq
F(Q) = \frac{3 j_1(qr)}{ qr } \; e^{-\frac{1}{2} (qs)^2},
\eeq
in units where $\hbar$ and $c$ are set to 1. $q = \sqrt{2Qm_{\rm N}}$, $s=1$ fm, $R = 1.2 A^{1/3}$ fm, $r = \sqrt{R^2 - 5s^2}$, and $j_1$ is the spherical bessel function. $F^2(Q) \approx 1$ for small $Q$, and falls off at large $Q$. We will also write $\sigma_0$ in terms of the scattering cross section with a proton or neutron $\sigma_{\rm p}$:
\beq
\sigma_0 = \sigma_{\rm p} \, A^2 \left ( \frac{m_{\rm R}}{m_{\rm R,p}} \right )^2.
\eeq
$A$ is the atomic mass number and $m_{\rm R,p} = m_\chi m_{\rm p} / (m_\chi + m_{\rm p})$ is the WIMP-proton reduced mass (we ignore the neutron-proton mass difference). We only consider spin-independent elastic scattering here. The recoil rate is
\beq
\frac{dR}{dQ}(t,Q) = \frac{\rho_\chi \sigma_{\rm p} A^2 }{2 m_\chi m^2_{\rm R,p}} F^2(Q) \; T(t,Q).
\label{rec}
\eeq
$T$ is the mean inverse speed
\beq
T = \int_{v_{\rm min}(Q)}^\infty   dv \;  \frac{f(v)}{v} ,
\label{mean_inverse_speed}
\eeq
which is time dependent due to the earth's motion about the sun. It is this term that leads to the annual modulation in recoil energy.

In order to apply Eq (\ref{rec}) to the DAMA experiment, we need to take into account (i) only a fraction of the deposited energy is detected and (ii) the target makes use of 2 elements, namely Na and I. The fraction of the recoil energy detected by the DAMA experiment is called the quenching factor which we label by q.f.(X) where X could be Na or I. In order to write Eq (\ref{rec}) in terms of the detected energy $Q_{\rm det}$, we make use of the equality of the total number of events
\beq
\frac{dR}{dQ} \, \Delta Q = \frac{dR}{dQ_{\rm det}} \, \Delta Q_{\rm det},
\eeq
and therefore,
\beq
\frac{dR}{dQ_{\rm det}}(X) = \frac{dR}{dQ} \, \left | \frac{\Delta Q}{\Delta Q_{\rm det}} \right | = \frac{1}{{\rm q.f.} (X)} \, \frac{dR}{dQ}.
\eeq
We can now write down the complete formula for the recoil rate:
\beq
\frac{dR}{dQ_{\rm det}} = \frac{A_{\rm Na}}{A_{\rm Na} + A_{\rm I}} \, \frac{dR}{dQ_{\rm det}}({\mathrm {Na}}) + \frac{A_{\rm I}}{A_{\rm Na} + A_{\rm I}} \, \frac{dR}{dQ_{\rm det}}({\mathrm I}).
\label{recoil_rate}
\eeq
We express our energies in terms of $Q_{\rm det}$ since this is the quantity measured by the experiment. We use the unit ``keV electron-equivalent (keV$_{\rm ee}$)'' to indicate that $Q_{\rm det}$ is being measured.

Since the DAMA experiment uses a crystalline detector, it is possible for an ion or recoiling nucleus moving parallel to the crystal axes to penetrate deep into the material. Such an ion is said to be channeled. Channeled ions transfer their energy primarily to electrons, leading to a near unity quenching factor \cite{chan}, i.e. $Q = Q_{\rm det}$. When the channeled fraction is known, the effect can be included in the calculation of the total recoil rate Eq (\ref{recoil_rate}).

\section{The self similar infall halo model}

A galactic halo is said to be self similar if its time evolution is such that the halo remains identical to itself except for an overall rescaling of its phase space density, and its size in spatial and velocity dimensions, by time dependent factors \cite{duffy}. Under the assumption of self similar evolution with the added assumption of spherical symmetry, Fillmore and Goldreich \cite{Fillmore} and Bertschinger \cite{bertschinger} described the properties of galactic halos. This model was modified to include angular momentum by Sikivie, Tkachev, and Wang \cite{sikivie1,sikivie2}.  For a recent review of the self similar model of the Milky Way halo, see \cite{duffy}. The self similar infall model predicts approximately flat rotation velocities far from the galactic center, in agreement with observations. The model is also consistent with the existence of a ``core radius'' observed in many galaxies. This is because most of the dark matter particles have angular momentum relative to the halo center and do not reach the central region, resulting in a depletion of dark matter particles relative to the spherically symmetric scenario \cite{duffy}.

\subsection{Discrete flows and caustics.}

A prominent feature of the self similar model is the existence of cold discrete flows and dark matter caustics. Note that the existence of discrete flows and caustics is not a consequence of self similarity, but rather a consequence of Liouville's theorem. In \cite{sikivie_ipser,arvi1}, it was argued that discrete flows and caustics should be a natural consequence of cold, collisionless matter. Each infall-outfall of dark matter produces an inner caustic and an outer caustic.  Outer caustics are fold catastrophes that occur at the outer turnaround radii of particles, and appear as thin spherical shells surrounding galaxies. Their location is determined by the energy of the particles. Inner caustics occur near the inner turnaround radii of particles. Their location is determined by the magnitude of angular momentum, and their geometry is determined by the spatial distribution of the dark matter angular momentum field. For the special case of dark matter particles carrying a net rotation aligned with that of the baryons, the inner caustics are made up of elliptic umbilic catastrophes that resemble rings in the galactic plane \cite{sikivie_crs,arvi2}. Possible observational evidence for such ring caustics and for self-similarity of galaxies was found by \cite{kinney_sikivie} who examined the rotation curves of spiral galaxies.  The existence of caustics is relevant to the DAMA experiment since the velocity distribution in the vicinity of a dark matter inner caustic is dominated by the cold flow forming the caustic. In \cite{arvi3}, the effect of the dominant flow of the self similar model on the annual modulation signature was calculated. The effect of other cold streams on the recoil rate has been discussed by several authors (see for example \cite{savage,streams1,streams2,streams3,streams4}). 

Table 1 (extracted from Table 1 of \cite{ling}) describes the first 40 flows in the self similar infall model of the Milky Way halo. The first column is the fractional density contribution of each flow. The table is arranged in descending order of the flow density fraction. The dominant flow is assigned a fraction $\xi$. This flow is dominant because the associated inner caustic is close to the earth's location \cite{ling, caustics_evidence} (In \cite{ling}, a value of $\xi=0.733$ is adopted). We allow $\xi$ to be variable since the value of $\xi$ determines the peak of the annual modulation. $\xi$ is also very sensitive to the location of the closest inner caustic.  The second and fourth columns give the maximum and minimum flow speeds relative to the earth. The third and fifth columns specify the time when the maximum and minimum occur respectively.
\begin{table} [!h] {The first 40 flows of the self similar infall model \\ 
       (from Table 1. of \cite{ling}) } \\
\vspace{0.1in}
\begin{tabular}  {|c |c |l |c |l |   @{  \hspace{0.2in}}  |c |c |l |c |l |} 
   \hline 

$\rho_i / \rho$  &  $v_{\rm max}$   &  $t_{\rm max}$  &  $v_{\rm min}$   &  $t_{\rm min}$   &  $\rho_i / \rho$  &  $v_{\rm max}$   &  $t_{\rm max}$  &  $v_{\rm min}$   &  $t_{\rm min}$  \\ 
        &      (km/s)  &      &  (km/s)   &      &     &   (km/s)   &       &   (km/s)     &  \\ \hline
$\xi$  & 273  &  Nov 5 &  234   &  May 7  & 0.0129(1-$\xi$)  & 543   & Dec 21 &  491   &  June 22  \\ \hline  
0.2427(1-$\xi$)  & 279  &  Jan 22 &  244   &  Jul 24  & 0.0113(1-$\xi$)  & 360  &  Sep 18 &  300   &  Mar 20  \\ \hline 
0.1052(1-$\xi$)  & 326  &  Feb 25 &  276   &  Aug 26  &  0.0113(1-$\xi$)  & 372  &  Sep 9 &  316   &  Mar 10  \\ \hline 
0.0663(1-$\xi$)  & 358  &  Mar 8 &  302   &  Sep 6  & 0.0113(1-$\xi$)  & 374  &  Sep 10 &  317   &  Mar 12  \\ \hline 
0.0550(1-$\xi$)  & 311  &  Oct 7 &  257   &  Apr 8  & 0.0113(1-$\xi$)  & 393  &  Mar 25 &  334   &  Sep 24  \\ \hline 
0.0550(1-$\xi$)  & 355  &  Jun 24 &  325   &  Dec 23  & 0.0113(1-$\xi$)  & 394  &  Mar 24 &  335   &  Sep 22  \\ \hline
0.0550(1-$\xi$)  & 382  &  Dec 19 &  322   &  Jun 19  & 0.0097(1-$\xi$)  & 370  &  Sep 6 &  316   &  Mar 8  \\ \hline 
0.0324(1-$\xi$)  & 365  &  Mar 11 &  307   &  Sep 10  & 0.0097(1-$\xi$)  & 372  &  Sep 7 &  317   &  Mar 9  \\ \hline 
0.0243(1-$\xi$)  & 380  &  Mar 16 &  321   &  Sep 15  & 0.0097(1-$\xi$)  & 391  &  Mar 27 &  333   &  Sep 26  \\ \hline 
0.0227(1-$\xi$)  & 444  &  Jun 21 &  400   &  Dec 20  & 0.0097(1-$\xi$)  & 392  &  Mar 26 &  334   &  Sep 25  \\ \hline 
0.0227(1-$\xi$)  & 464  &  Dec 20 &  408   &  Jun 21  & 0.0081(1-$\xi$)  & 359  &  Sep 4 &  306   &  Mar 5  \\ \hline 
0.0210(1-$\xi$)  & 340  &  Sep 27 &  281   &  Mar 28  & 0.0081(1-$\xi$)  & 361  &  Sep 4 &  308   &  Mar 6  \\ \hline 
0.0178(1-$\xi$)  & 346  &  Sep 23 &  286   &  Mar 25  & 0.0081(1-$\xi$)  & 362  &  Sep 5 &  309   &  Mar 7  \\ \hline 
0.0162(1-$\xi$)  & 376  &  Sep 15 &  317   &  Mar 16  & 0.0081(1-$\xi$)  & 379  &  Mar 29 &  322   &  Sep 28  \\ \hline 
0.0162(1-$\xi$)  & 396  &  Mar 20 &  337   &  Sep 18  & 0.0081(1-$\xi$)  & 382  &  Mar 29 &  324   &  Sep 27  \\ \hline 
0.0146(1-$\xi$)  & 373  &  Sep 13 &  315   &  Mar 15  & 0.0081(1-$\xi$)  & 383  &  Mar 28 &  325   &  Sep 26  \\ \hline 
0.0146(1-$\xi$)  & 394  &  Mar 21 &  334   &  Sep 20  & 0.0065(1-$\xi$)  & 354  &  Sep 2 &  302   &  Mar 4  \\ \hline 
0.0129(1-$\xi$)  & 376  &  Sep 11 &  319   &  Mar 13  & 0.0065(1-$\xi$)  & 374  &  Mar 31 &  317   &  Sep 29  \\ \hline 
0.0129(1-$\xi$)  & 397  &  Mar 23 &  338   &  Sep 21  & 0.0049(1-$\xi$)  & 635  &  Jun 18 &  579   &  Dec 18  \\ \hline 
0.0129(1-$\xi$)  & 529  &  Jun 19 &  477   &  Dec 19 & 0.0049(1-$\xi$)  & 644  &  Dec 23 &  598   &  Jun 23  \\ \hline 
   \hline   
\end{tabular}

\vspace{0.1in}
\caption{  The first 40 flows and their associated density fractions, from \cite{ling}, in descending order of density contribution. The first column gives the fraction of the total density contributed by each flow. The dominant flow is assigned a density fraction $\xi$.  $v_{\rm max}$ and $v_{\rm min}$ are the maximum and minimum flow speeds relative to the earth, seen at times $t_{\rm max}$ and $t_{\rm min}$. The recoil energy maximum observed by DAMA in the $2-6$ ${\rm keV_{ee}}$ range during May 17 $< t < $ June 2 is obtained for $0.62 > \xi > 0.37$. The mean DAMA best fit maximum on May 25 is obtained for $\xi = 0.47$, while a peak on June 2 would correspond to a density fraction $\xi = 0.37$. The flow densities published in \cite{ling} are obtained by setting $\xi = 0.733$ and multiplying by $231.8 \times 10^{-26}$ gm/cm$^3$. Note that the flow velocities in \cite{ling} are related to the velocities used here by the transformation $\hat x \rightarrow -\hat x, \hat y \rightarrow \hat y, \hat z \rightarrow -\hat z$. } 
\end{table}

\subsection{Annual modulation.}
For a series of cold flows (i.e. ignoring the velocity dispersion, a valid assumption for WIMPs), the velocity distribution of WIMPs (relative to the halo) is:
\beq
f_{\rm flows}(\vec v) = \sum_i \xi_{\rm flows,i} \,  \delta(\vec v - \vec v_{f,i}),
\label{flows}
\eeq
where $\xi_{\rm flows,i}$ represents the contribution of the $i^{\rm th}$ flow to the local dark matter density.  The mean inverse speed $T(t,Q)$ can then be easily calculated for a flow $v_f$:
\bea
T(t,Q) &=& \frac{1}{\left | \vec v_{f\oplus}(t) \right |} \theta [ |\vec v_{f\oplus}(t)| - v_{\rm min}(Q) ] \n
 &\approx& \frac{1}{v_{f \odot}} \; \left[ 1 + \frac{v_\oplus}{v_{f \odot}} \; \hat v_\oplus (t) \cdot \hat v_{f \odot} \right ] \; \theta \left[ v_{f \odot} - v_\oplus \, (\hat v_\oplus(t) \cdot \hat v_{f\odot}) - v_{\rm min}(Q) \right ],
\label{T_flow}
\eea
where $v_{f\oplus}$ and $v_{f\odot}$ are the flow velocities relative to the earth and sun respectively, and $\theta$ denotes the unit step function. $v_\oplus(t)$ is the velocity of the earth about the sun, and $v_\odot$ is the velocity of the sun about the halo center.

\begin{figure}[!h]
\begin{center}
\scalebox{0.65}{\includegraphics{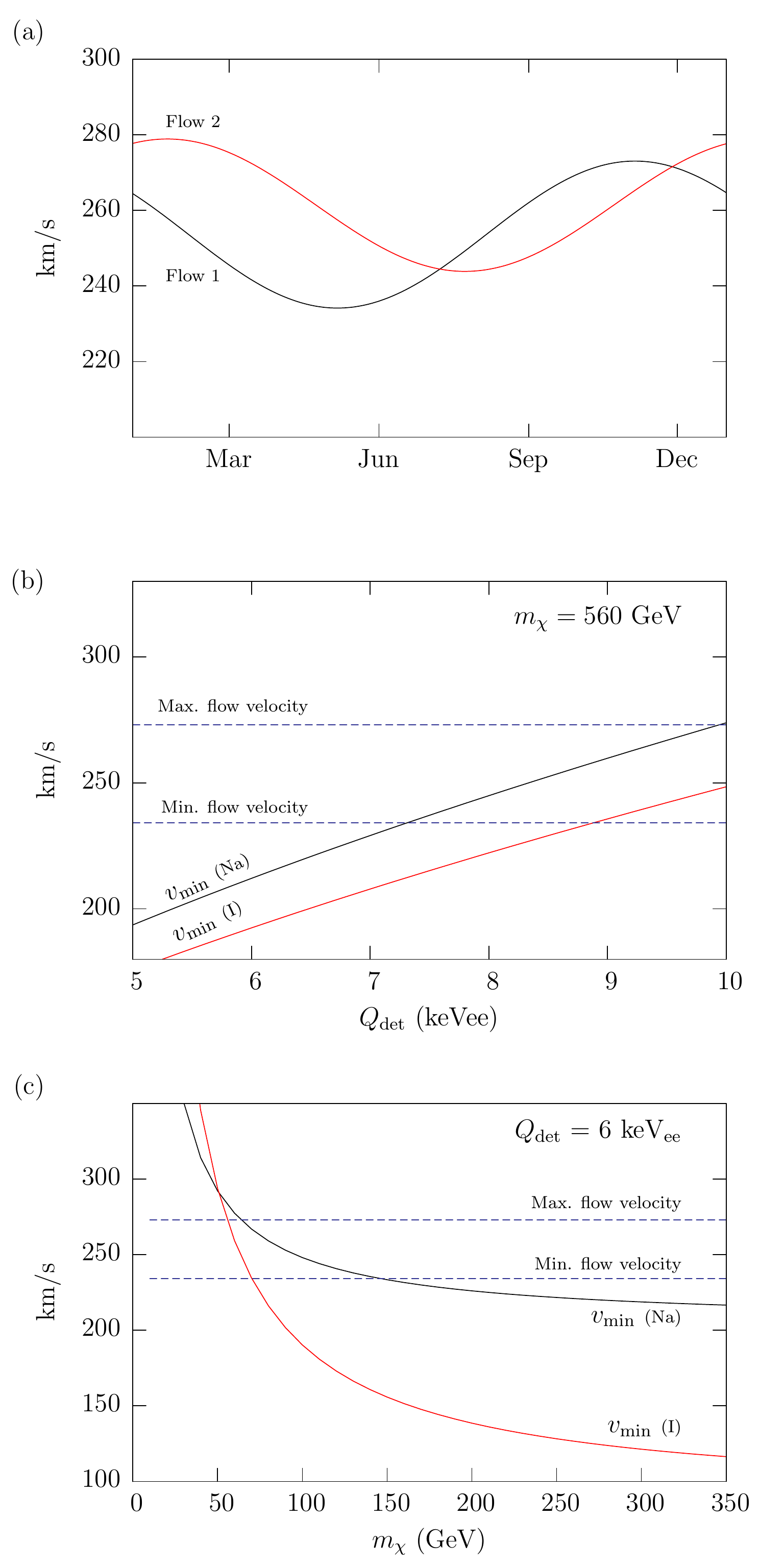}}
\end{center}
\caption{  (a) shows the first 2 flows of the self similar model. (b) shows $v_{\rm min}$ for Na and I as a function of $Q_{\rm det}$ for a WIMP mass $m_\chi$ = 560 GeV, for quenched recoils. Also shown are the maximum and minimum velocities for Flow 1. (c) shows $v_{\rm min}$ for Na and I for quenched recoils, for different values of $m_\chi$, for fixed $Q_{\rm det} = 6$ keV$_{\rm ee}$.    \label{fig2} }
\end{figure}

Fig. \ref{fig2}(a) shows the first 2 flows. Flow 1 is smallest in May and peaks in November. Flow 2 is smallest in July and peaks in January.  Since Flow 1 is the dominant flow, it is instructive to look at this flow in detail.  Let us choose a co-ordinate system in which the $+\hat x$ axis points towards the Galactic center, the $+\hat y$ axis points in the direction of Galactic rotation, and the $+\hat z$ axis points towards the north Galactic pole. In these co-ordinates, relative to the sun, Flow 1 has velocity \cite{ling} (but note that the co-ordinate system used in \cite{ling} is different from ours):
\beq
\vec v_{1, \odot} = 253.6 \, \textrm{km/s} \; \left( 0.3549 \, \hat x + 0.9345 \, \hat y - 0.0276 \, \hat z \right ) 
\label{flow_vel}
\eeq
The velocities of the sun (about the halo center) and the earth (about the sun) in these co-ordinates are respectively (see for example \cite{gelmini,streams2,savage}, and references therein):
\bea
\vec v_\odot &=& 233.3 \, \textrm{km/s} \; \left[ 0.0429 \, \hat x + 0.9986 \, \hat y + 0.0300 \, \hat z \right ] \n
\vec v_\oplus (t) &=& 29.8 \, \textrm{km/s} \; \left [ (0.9931 \cos\phi - 0.0670 \sin\phi) \, \hat x \right. \n
&+& (\left. 0.1170 \cos\phi + 0.4927 \sin\phi) \, \hat y - (0.0103 \cos\phi + 0.8676 \sin\phi) \, \hat z \right ], 
\label{earth}
\eea
where the angle $\phi(t) = 2 \pi \; (t - \textrm{March} \; 21)/365$. We note that earth's velocity about the sun is most closely aligned with the sun's velocity about the halo center when $\phi = 71 ^\circ$, which occurs around June 2. The two velocity vectors are most misaligned when $\phi = 251 ^\circ$ which occurs six months later, around Nov 30. Flow 1 has speed relative to the earth:
\beq
\left | \vec v_{1,\oplus}(t) \right | \approx v_{1,\odot} - v_\oplus \; \left [ \hat v_\oplus(t) \cdot \hat v_{1,\odot} \right ],
\eeq
which is largest when $\phi = 225^\circ$ (around Nov 5) and smallest when $\phi = 45 ^\circ$ (around May 7). 

Fig. \ref{fig2}(b) shows $v_{\rm min}$ for quenched recoils, for Na and I as a function of $Q_{\rm det}$, for an assumed $m_\chi = 560$ GeV (the value of $m_\chi$ is motivated by the data, as shown in the next section). Also shown are the minimum and maximum velocities for Flow 1, which occur on May 7 and Nov 5 respectively. For $Q_{\rm det} < 7.4$ keV$_{\rm ee}$, the flow is visible to the detector at all times of the year. We therefore expect a sinusoidal variation of $dR/dQ$ with a peak in May (for this flow). For energies 7.4 keV$_{\rm ee} < Q_{\rm det} <$ 8.9 keV$_{\rm ee}$, the flow in invisible to Na during parts of the year. For energies 8.9 keV$_{\rm ee} < Q_{\rm det} <$ 10 keV$_{\rm ee}$, the flow is invisible to both Na and I during parts of the year, resulting in pronounced non-sinusoidal behavior. In the 10 keV$_{\rm ee} < Q_{\rm det} <$ 12.1 keV$_{\rm ee}$ range, the flow is completely invisible to Na and is seen by I only during some parts of the year near the flow maximum in November. For $Q_{\rm det} > 12.1$ keV$_{\rm ee}$, the flow is invisible to both Na and I, resulting in no recoils for this particular flow. Fig. \ref{fig2}(c) shows $v_{\rm min}$ as a function of WIMP mass, for $Q_{\rm det} = 6$ keV$_{\rm ee}$, for quenched recoils.  For masses $m_\chi > 150$ GeV, the flow is visible to both Na and I during all parts of the year. Thus for sinusoidal variation of $dR/dQ$ at 6 keV$_{\rm ee}$, we must have $m_\chi > 150$ GeV. We will see in the next section that more stringent bounds can be obtained. The recoil rate $dR/dQ \propto T(t,Q) \propto 1 / \left |\vec v_{1,\oplus} \right |$ provided the flow is visible to the detector at energy $Q$.

It is interesting to contrast the self similar model with the Maxwellian. A Maxwellian halo is described by the distribution:
\beq
f_{\rm max}(\vec v_{wh}) = \frac{\exp \left[ - (\vec v_{wh} / v_0)^2 \right ]}{\pi^{3/2} \; v^3_0 },
\label{max}
\eeq
where we have ignored the effect of the finite escape velocity. The subscript $wh$ stands for ``WIMP-halo'' and indicates that the velocities are measured relative to the halo. Expressed relative to the detector, the (1-dimensional) speed distribution becomes:
\beq
f(v) = \frac{v}{\sqrt{\pi} v_0 v_{eh}} \; \left[ e^{-\left( \frac{v-v_{eh}}{v_0} \right )^2 } - e^{-\left( \frac{v+v_{eh}}{v_0} \right )^2 } \right ],
\eeq
implying a mean inverse speed
\beq
T_{\rm max}(t,Q) = \frac{1}{2 v_{eh}(t)} \; \left[ \textrm{erf} \left \{ \frac{v_{\rm min}(Q) + v_{eh}(t)}{v_0} \right \} - \textrm{erf} \left \{ \frac{v_{\rm min}(Q) - v_{eh}(t)}{v_0} \right \} \right ],
\label{Tmax}
\eeq
where $v_{eh}(t) = | \vec v_\odot + \vec v_\oplus(t) |$.   
It is instructive to compute the angular dependence of the flux of dark matter particles on earth for the self similar infall model, and contrast it with the prediction of the isothermal halo, as done in \cite{gelmini}. Since the streams of the self similar model have a net velocity relative to the halo center, they do not all arrive from the direction of the sun's motion. Some streams arrive in directions above and below the galactic plane, but the densest streams (and in particular, the dominant, or big flow) are restricted to the galactic plane, in a direction nearly opposite to that of the sun's motion. In contrast, for the isothermal halo, the WIMP particles have no average velocity relative to the halo center, and the WIMP wind is due to the motion of the sun (and earth), implying a much larger flux in the direction of the sun's motion compared to the flux in the opposite direction. 

Also of note is the energy spectrum of recoils. In the self similar model, the velocity distribution is discrete. This means that at a fixed energy, a stream may or may not be visible, and the number of streams that contribute to the signal decreases as the recoil energy is increased. The mean inverse speed $T$ given by Eq. \ref{mean_inverse_speed} is therefore a series of steps, for the self similar model. The height of each step $\propto 1/v$, where $v$ is the speed of the flow under consideration (relative to the detector), and is therefore largest at times of the year when the flow speed relative to the earth is the smallest. The edge of the step on the other hand $\propto v^2$, and is largest for the highest velocity flows. The modulation of the edge of each step is twice the modulation of the height, and has opposite phase. The net recoil rate integrated over energies depends not only on the flow speeds and densities, but also on the energy dependence of the cross section. As $F^2(Q) \ll 1$ at high energies, the DAMA experiment is sensitive to recoils only for $Q_{\rm det} < 6$ keV$_{\rm ee}$, and relatively low energy WIMP recoils contribute more than high energy recoils. 

 For the self similar model, $T$ peaks in May/June (depending on the value of $\xi$) for large WIMP masses. For a fixed WIMP mass, $T$ peaks in May/June at small energies, becoming non-sinusoidal, and possibly reversing phase at larger energies. In contrast, for the isothermal halo with a Maxwellian velocity distribution,  $T$ measured at a fixed recoil energy peaks in June for sufficiently small WIMP masses, but reverses phase and peaks in November for larger WIMP masses, as can be verified by expanding Eq. \ref{Tmax} in a Taylor series \cite{mass_phase}. Conversely, for a fixed WIMP mass, $T$ peaks in June at large energies, but reverses phase, peaking in November at very small energies. Given the very different nature of these two halo models, it is certainly surprising that the DAMA results may be fit to either model as we shall see in the following section (when the WIMP mass is not fixed by an independent measurement).

\section{Results}

We now compare the prediction of the self similar halo model with the DAMA results and obtain best fit values of mass and cross section. We set the dark matter density at our location $\rho_\chi = 0.3$ GeV/cm$^3$. We assume energy independent values for the quenching factor q.f. = 0.3 for Na and q.f. = 0.09 for I. We consider only the spin-independent cross section for WIMPs scattering elastically off a proton (or neutron) $\sigma_{\rm p}$.  The fraction of channeled recoils for Na and I as a function of energy has been calculated experimentally by \cite{chan}. For a recent theoretical treatment, see \cite{gondolo_chan} . Here we use the fit obtained by \cite{savage2} to Figure 4 of \cite{chan}, for the channeling fractions for Na and I. 

We use Table 1 of \cite{ling} (summarized in Table 1 here) to obtain the flow velocities and densities. We use Table 2 published in \cite{dama2} which gives the observed peak position in the recoil energy spectrum for different energies, in order to determine $\xi$. The amplitude of the recoil spectrum is presented in \cite{dama2}. We fit the recoil rate at different energies to the observed amplitude of the annual modulation using Table 3 published in \cite{savage2} which was extracted from Fig. 9 of \cite{dama2}. We perform a minimum $\chi^2$ analysis using 36 energy bins and 2 fitting parameters ($m_\chi$ and $\sigma_{\rm p}$). The contours are obtained by plotting curves of constant $\chi^2 = \chi^2_{\rm min} + \Delta\chi^2$. $\Delta\chi^2$ is obtained by setting the area under the $\chi^2$ distribution equal to the required confidence level ({\textrm {C.L.}}), with $n$ equal to the number of fitting parameters (see for e.g. \cite{savage2}, \cite{fairbairn}):
\beq
\frac{1}{2^{\frac{n}{2}} \Gamma(\frac{n}{2})} \int_0^{\Delta\chi^2} d\chi^2 \left[ \chi^2 \right ] ^{\frac{n}{2}-1} \; e^{-\frac{1}{2}\chi^2} = {\textrm {C.L.}}
\eeq
For $n=2$, this simplifies to $\Delta\chi^2 = -2 \, \log (1-{\textrm {C.L.}})$, and we find $\Delta\chi^2$ = 2.16, 5.99, and 11.62 for 66\%, 95\%, and 99.7\% confidence respectively.

\subsection{Fitting the location of the maximum recoil rate.}

The maximum in the recoil spectrum measured by the DAMA experiment in the energy range $2-6$ keV$_{\rm ee}$ is $t_{\rm max} = 144$ (May 25) $ \pm 8$ days. We vary $\xi$ to fit the DAMA phase. We find that $\xi = 0.47$ fits $t_{\rm max} = 144$, implying that the dominant flow contributes 47\% of the local dark matter density.  Such a large contribution implies the existence of a nearby dark matter inner caustic (in \cite{caustics_evidence}, an observation of such a caustic is claimed). The observed maximum of $144 \pm 8$ days is obtained for $0.62 > \xi > 0.37$, with $\xi=0.62$ corresponding to a peak on May 17, while $\xi=0.37$ giving a peak on June 2. We note that the DAMA maximum picks Flow 1 as the dominant flow, while in \cite{ling}, the dominant flow is either Flow 1 or Flow 2. $\xi$ is set to 0.47 for all our results.

\subsection{Best fit parameters.}

We fit our 2 free parameters $m_\chi$ and $\sigma_{\rm p}$ by minimizing 
\beq
\chi^2 = \sum_{i=1}^{36} \left [ \frac{A_{\rm data,i} - A_{\rm model,i}(\sigma_{\rm p}, m_\chi)}{\sigma_i} \right ]^2,
\label{chi_squared}
\eeq
where the sum is over energy bins. $A_{\rm data,i}$ is the measured amplitude for energy bin $i$ and $A_{\rm model,i}(\sigma_{\rm p}, m_\chi)$ is the predicted amplitude for energy bin $i$ for the assumed $m_\chi$ and $\sigma_{\rm p}$. $\sigma_i$ is the uncertainty in the measurement of $A_i$. We compute the amplitude $A_{\rm model}$ as 
\beq
A_{\rm model} = \frac{1}{2} \left[ dR/dQ_{\rm det} (t_{\rm max}) - dR/dQ_{\rm det} (t_{\rm min}) \right ], 
\eeq
in the energy region where $A_{\rm model}$ is sinusoidal. For the self similar model, we use the DAMA best fit value of $t_{\rm max}$ = 144. $t_{\rm min}$ is set equal to 327. When a comparison with the isothermal halo is made, we use Eq. \ref{max} with $v_0 = 220$ km/s, $t_{\rm max}$ = 152, and $t_{\rm min} = 335$.
\begin{figure}[!h]
\begin{center}
\scalebox{0.7}{\includegraphics{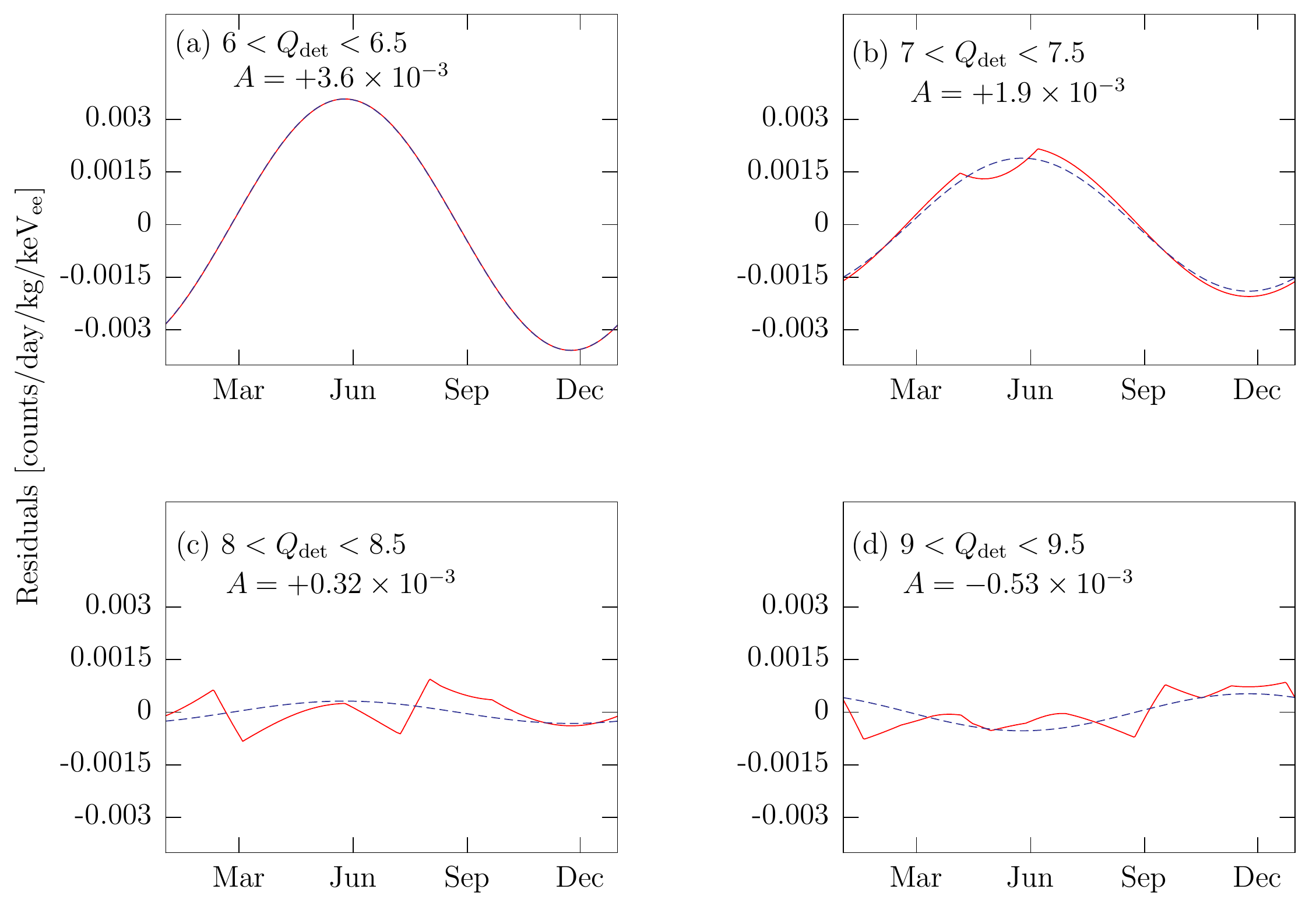}}
\end{center}
\caption{   Recoil rate (mean subtracted) expected for the self similar infall halo model at different times of the year, for 4 different energy bins. The rate is a sinusoidal function of time when the flow velocity exceeds $v_{\rm min}$ at all times of the year as in (a). For large energies [(c) and (d)], the flow is only visible to the detector during parts of the year, resulting in a non-sinusoidal pattern and phase reversal in (d). Ion channeling is included.\label{fig3} }
\end{figure}

Fig. \ref{fig3} shows the expected recoil rate in the self similar infall model for 4 different energy bins.  For energies $Q_{\rm det} < 6$ keV$_{\rm ee}$, the recoil spectrum is qualitatively identical to (a). The expected sinusoidal variation with a maximum at $t=144$ and a minimum at $t=327$ is seen in (a). When the energy range is increased in (b), non-sinusoidal features start to appear since the flow velocity near $t=144$ is less than the minimum velocity $v_{\rm min}$ required to produce Na recoils at this energy (see Fig \ref{fig2}(b) for the dominant flow). As the energy is increased, $v_{\rm min}$ increases, leading to the non-sinusoidal shapes seen in (c) and (d). The amplitude is negative in (d) indicating a phase reversal. DAMA does measure negative amplitudes at high energies, in particular the measurement in the energy range $9.5-10$ keV$_{\rm ee}$ is statistically significant. Smaller error bars are required before the negative amplitudes measured by DAMA can be treated as a physical effect.  Negative amplitudes measured at high energies favor the self similar model (or a cold stream) and cannot be accommodated by the isothermal halo. The very small amplitudes seen in (c) and (d) due to the small value of $F^2(Q)$ make detection challenging.

\begin{figure}[!h]
\begin{center}
\scalebox{0.6}{\includegraphics{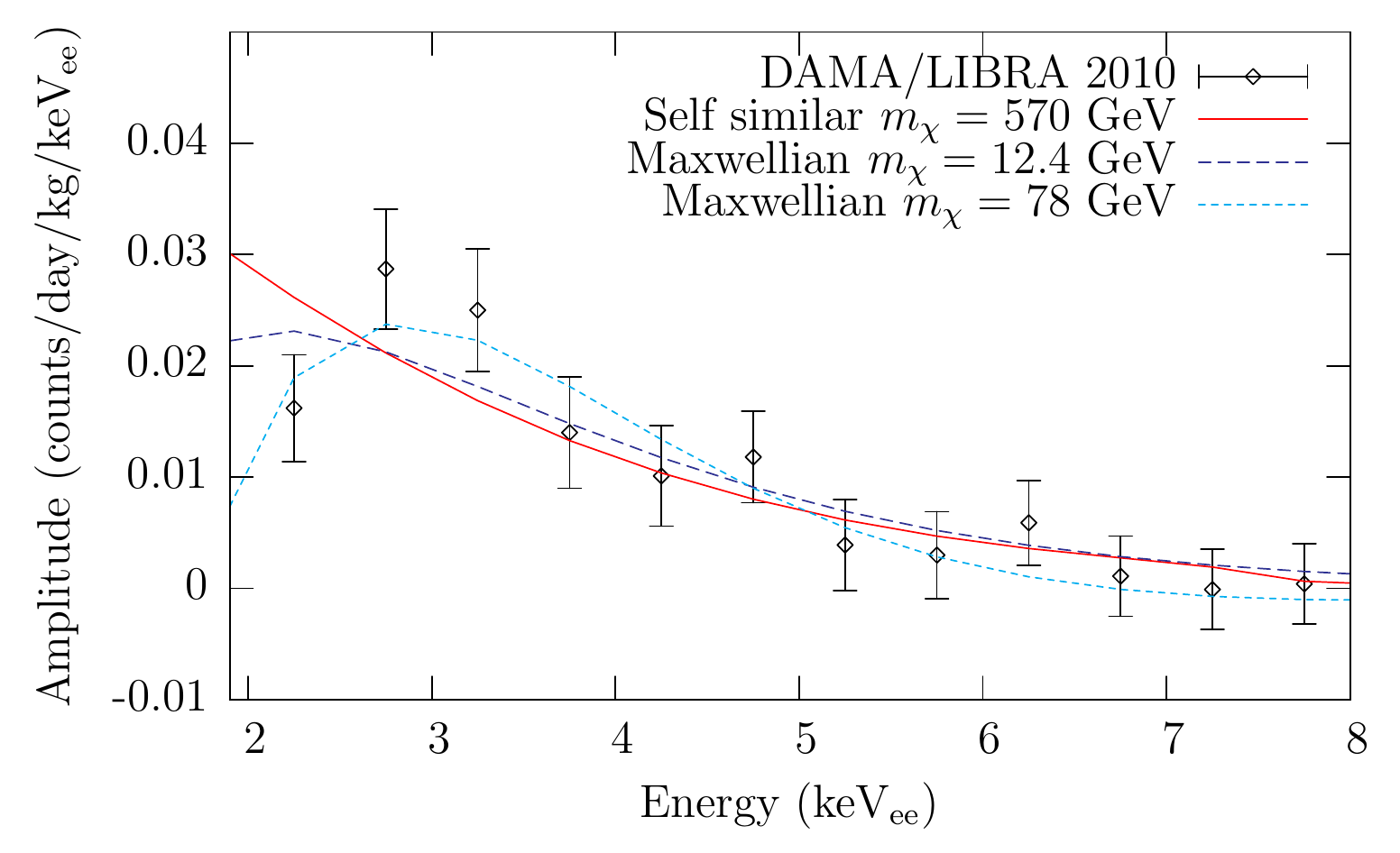}}
\end{center}
\caption{   Modulation amplitudes. Points are the DAMA/LIBRA measurements. The solid (red) line is the prediction of the self similar model, while the broken lines are drawn for the two Maxwellian models.  \label{fig4} }
\end{figure}

\begin{figure}[!h]
\begin{center}
\scalebox{0.8}{\includegraphics{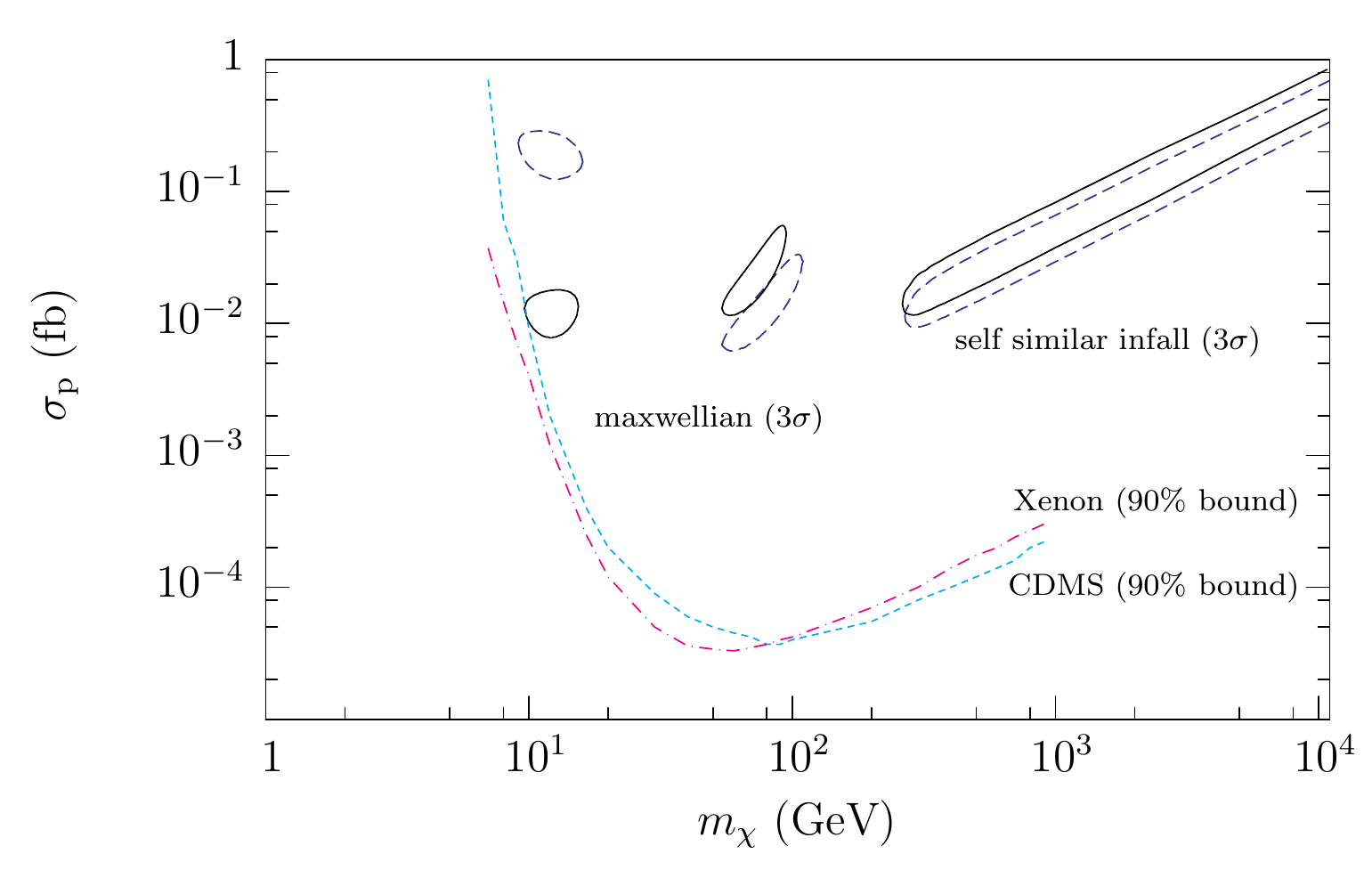}}
\end{center}
\caption{  Allowed regions in parameter space for elastic spin-independent scattering. The contours with solid lines include the effect of ion channeling, while the contours with broken lines do not. Shown are $3\sigma$ results. The 3$\sigma$ contour for the self similar model is not closed at the high mass end because for $m_\chi$ much greater than the mass of an Iodine nucleus, $v_{\rm min}$ is nearly independent of $m_\chi$. The $1\sigma$ contour (not shown) is bounded between 330 GeV and 3.3 TeV. The allowed regions are ruled out by the CDMS and Xenon bounds (using data from \cite{xenon}). Also shown are the allowed regions for the Maxwellian halo of Eq. \ref{max}.  $\chi^2_{\rm min}$ is found to be 31.31/34 dof at $m_\chi = 570$ GeV for the self similar model. For the Maxwellian we find $\chi^2_{\rm min}$ = 30.6/34 dof at $m_\chi = 12$ GeV, and $\chi^2_{\rm min}$ = 26.41/34 dof at $m_\chi = 78$ GeV. \label{fig5} }
\end{figure}

Fig. \ref{fig4} shows the modulation amplitudes measured by the DAMA collaboration (open diamonds with error bars) at different energy bins. The prediction of the self similar model ($m_\chi$ = 570 GeV) is shown by the solid line, while the amplitudes expected for the Maxwellian models ($m_\chi$ = 78 GeV and $m_\chi$ = 12.4 GeV) are shown by broken lines.

Fig. \ref{fig5} shows the 99.7\% confidence contours in the mass-cross section plane for elastic spin-independent scattering, for the self similar infall model. The contours with solid lines include the effect of ion channeling while the contours with broken lines do not. The channeling effect is more important at lower WIMP masses.  Using Eq. (\ref{chi_squared}), the minimum value $\chi^2_{\rm min}$ is found to be $31.31/34$ dof when the effect of channeling is included, and  $34.93/34$ dof without channeling at $m_\chi \approx$ 570 GeV (dof stands for degrees of freedom).

The contour for the self similar model is not closed. This is due to the fact that for WIMP masses $m_\chi$ much greater than the mass of Iodine, the minimum velocity
\beq
v_{\rm min} = \sqrt{ \frac{Q m_{\rm N}}{2 m^2_{\rm R}} } \approx \sqrt{ \frac{Q}{2 m_{\rm N}} }
\eeq
becomes independent of WIMP mass $m_\chi$. As a result, the recoil rate only depends on one parameter $\sigma_{\rm p} / m_\chi$, and not on $\sigma_{\rm p}$ and $m_\chi$ separately. Fitting the data to the single variable $\sigma_p/m_\chi$, we find a best fit value of $0.059 \pm 0.014$ fb/TeV at the 95\% level, with a $\chi^2_{\rm min}$ of $35.43/35$ for the one parameter fit. The 66\% contour (not shown) is closed between WIMP masses 330 GeV and 3.3 TeV. Shown for comparison are the contours for the simple Maxwellian model of Eq. \ref{max}. The CDMS and Xenon100 exclusion contours are also shown (CDMS and Xenon data from Fig. 5 of \cite{xenon}).

\subsection{Adding a thermal component.}
Let us now modify our discrete sum over cold flows by adding a thermal component. This is similar to adding numerous flows that approximate a continuum to an experiment with finite energy resolution. We would naturally expect the innermost region of phase space to be unresolved to detectors while the outer region of phase space to be seen as a series of cold flows. We modify Eq. \ref{flows} as:
\beq
\rho f(\vec v) = \rho \, \left [ \xi_{\rm flows} f_{\rm flows} + (1-\xi_{\rm flows}) f_{\rm max} \right ] ,
\label{th}
\eeq
where $\xi_{\rm flows}$ is the contribution due of the 40 flows in Table 1, and $f_{\rm max}$ is given by Eq. \ref{max}.

\begin{figure}[!h]
\begin{center}
\scalebox{0.8}{\includegraphics{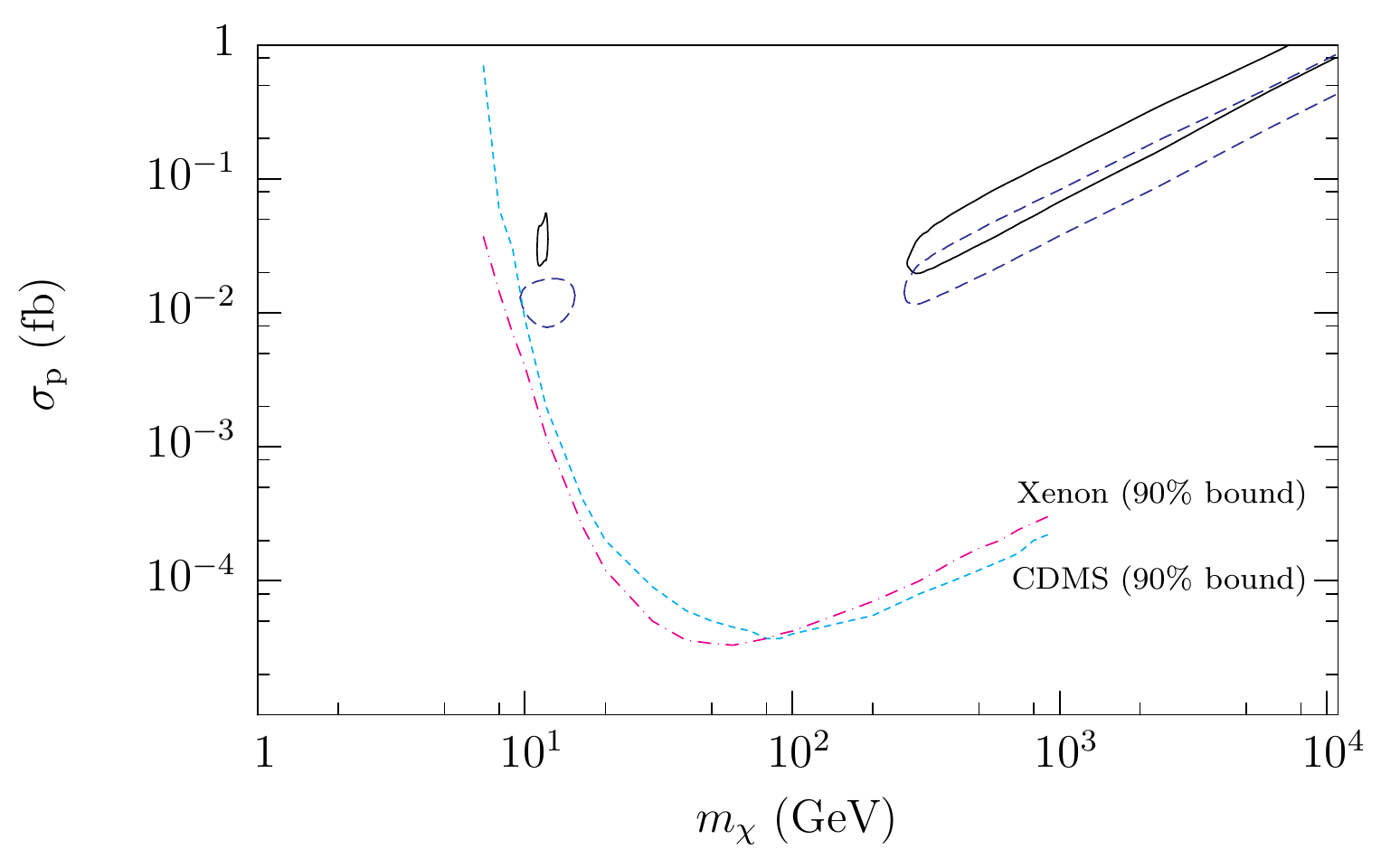}}
\end{center}
\caption{Effect of adding a thermal component. The 3$\sigma$ contours with solid lines are for the distribution of Eq (\ref{th}) with $\xi_{\rm flows}=0.7$. The allowed region becomes larger as $\xi_{\rm flows}$ is reduced. The contours with broken lines are for different distributions (self similar or Maxwellian). Ion channeling is included. \label{fig6} }
\end{figure}
 We choose $\xi_{\rm flows}$ = 0.7 so that the thermal component is comparable to the dominant flow. Fig. \ref{fig6} shows the allowed $3\sigma$ contours. Also shown (in dashed lines) are the contours for the Maxwellian halo at low masses, and the self similar infall model for high masses (Note that the contours with solid lines are for the same $f(\vec v)$ (Eq. \ref{th}), and the contours with dashed lines are for different models).  There exists a tiny region near $m_\chi = 12$ GeV which is affected by some of the high velocity flows, but not by the dominant flow. The $3\sigma$ contour is much smaller than for a pure Maxwellian halo, but becomes larger as $\xi_{\rm flows}$ is decreased. Also $\sigma_{\rm p}$ is about three times the value for a Maxwellian halo, as expected for $\xi_{\rm flows} = 0.7$. Thus the cold streams of the self similar model do not help in bringing the DAMA result in agreement with other experiments. It is however important to note that the self similar model is consistent with a small WIMP mass provided a thermal component exists. The other allowed region is at high WIMP masses $m_\chi > 250$ GeV, as with the case of pure self similar infall, with slightly larger $\sigma_{\rm p}$ to compensate for the smaller value of $\xi_{\rm flows} = 0.7$.  The $\chi^2_{\rm min}$ is found to be $30.64/34$ dof near $m_\chi=12$ GeV and $30.05/34$ dof near $m_\chi=570$ GeV. The $\chi^2$ function also has a minimum near $m_\chi=48$ GeV, but the large value ($\approx 61$) means it is disfavored at high significance.

\subsection{Comparison with the DAMA average plus unidentified background.}

\begin{figure}[!h]
\begin{center}
\scalebox{0.3}{\includegraphics{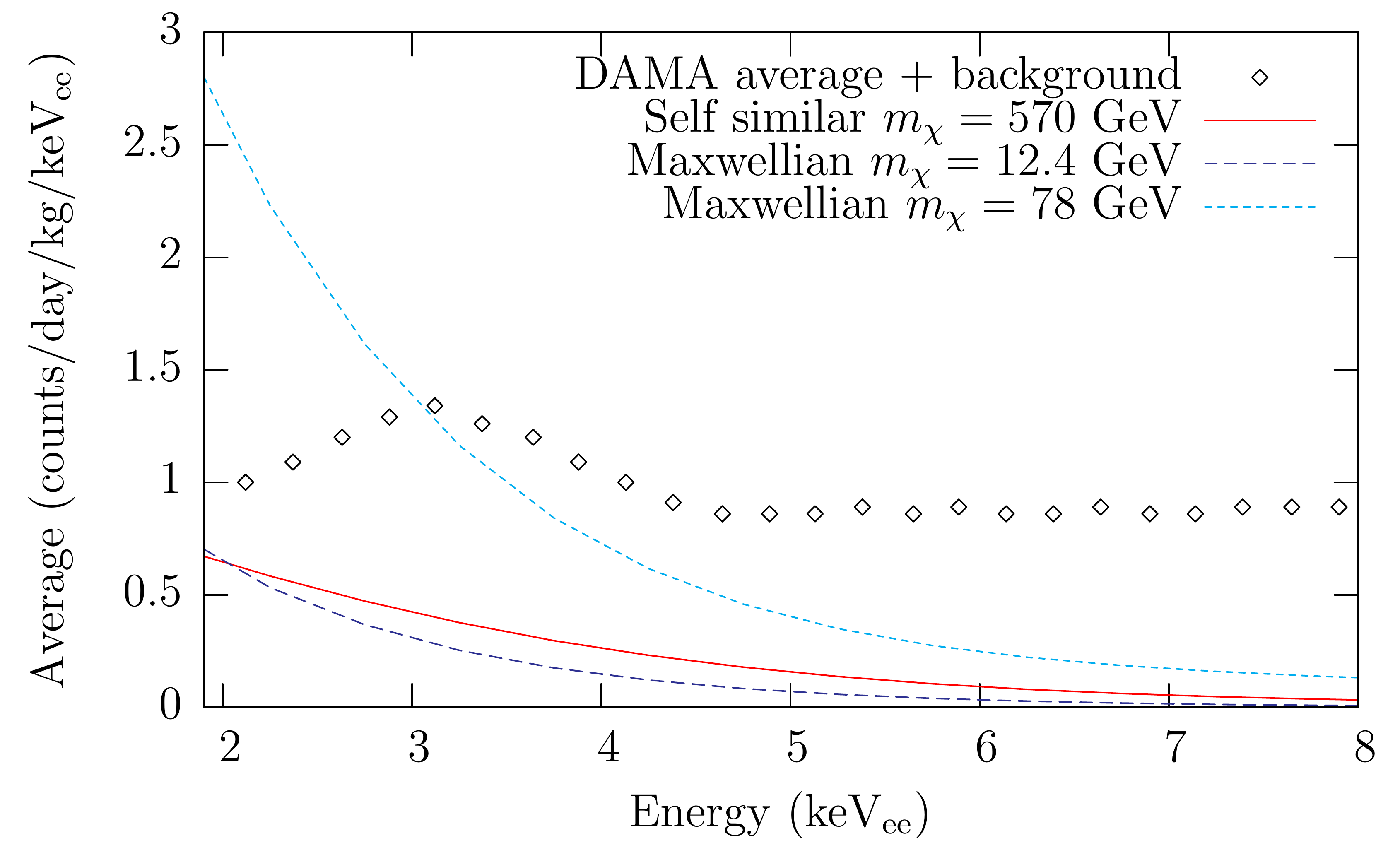}}
\end{center}
\caption{ Comparison with the measured average + background. The open diamonds (black) are the DAMA measurement of the average + background, extracted from Fig. 1 of \cite{dama2}. The solid line (red) is the expected average recoil rate for the self similar model. We see that the line remains below the measured average + background at all energies. The Maxwellian models are shown by broken lines for $m_\chi = 12.4$ GeV and $m_\chi = 78$ GeV.  \label{fig7} }
\end{figure}

The DAMA experiment measures only the modulation about the average recoil rate, not the average itself. This is because of the presence of a large background that does not modulate annually. Nevertheless, the DAMA collaboration has reported the sum of the average recoil rate and the unidentified background \cite{dama2}. It is an important check that the average recoil rate predicted by a particular model is less than the average plus background reported by DAMA. Figure \ref{fig7} shows the DAMA measurement (open diamonds) extracted from Fig. 1 of \cite{dama2} compared with the self similar model (solid line) and the two Maxwellian models (broken lines). We see that the prediction of the self similar model is everywhere below the average plus background reported by DAMA, and is thus consistent with the observations. The Maxwellian model with $m_\chi = 78$ GeV predicts an average rate that is too large in the first few energy bins. We will however not exclude that solution as our treatment of the isothermal halo is quite approximate. In \cite{fairbairn}, a more thorough analysis leads to conclusions similar to what we find here regarding the Maxwellian model with relatively large WIMP mass $\sim 80$ GeV. 

\section{Discussion.}

In this paper, we examined the DAMA annual modulation result in the context of the self similar infall halo model. We showed that the self similar model is in good agreement with the DAMA experiment for spin independent elastic scattering, with $\chi^2_{\rm min}$ per degree of freedom = 0.92(1.03) with(without) channeling, for WIMP masses exceeding 250 GeV at 99.7\% confidence. For large WIMP masses, the cross section-mass relation is approximately $\sigma_{\rm p} / m_\chi \approx$ 0.06(0.05) fb/TeV with(without) channeling. As in the case of the Maxwellian, the allowed region has been excluded by the CDMS and Xenon experiments. 

In Section II, we derived an expression for the expected recoil rate  assuming a spin-independent cross section and elastic scattering. We then discussed the self similar infall model in Section III. We examined the speed of the dominant flow at different times of the year, and compared it to $v_{\rm min}$ for both Na and I (Fig. \ref{fig2}). We then presented our results in Section IV. Fig. \ref{fig3} shows the modulation amplitude (mean subtracted) expected for the self similar model at four different energy bins. Fig. \ref{fig4} shows the best fit amplitudes compared to the DAMA/LIBRA measurement, while Fig. \ref{fig5} shows the allowed regions in parameter space with and without ion channeling taken into account. We then introduced a small thermal component and studied the effects (Fig. \ref{fig6}). With a thermal component, there are two allowed regions in parameter space. The first is near $m_\chi=12$ GeV due to the thermal component. This region is only slightly affected by the flows. The other is near $m_\chi$ = 570 GeV and is due to the cold flows. Finally, we verified (Fig. \ref{fig7}) that the time averaged recoil rate predicted by the self similar model is consistent with the measurement of the average plus background reported by the DAMA collaboration.

If the DAMA results are indeed correct, it is interesting to ask whether we can distinguish the different halo models. We have shown here that the self similar model has properties that distinguish it from the isothermal Maxwellian. Let us take a look at the differences:

\subsubsection{Negative amplitudes and measurements at low energies.}
As shown in Fig. \ref{fig3}, the recoil rate in the self similar infall model becomes non-sinusoidal at large energies, with the rate in November exceeding the rate in May/June. A convincing measurement of negative amplitudes (phase reversal) in the high energy bins would be consistent with the self similar infall model, but inconsistent with the isothermal Maxwellian halo. The DAMA result does include a statistically significant negative amplitude in the 9.5-10 keV$_{\rm ee}$ bin. However most measurements at high energies are smaller than the uncertainty, and are thus not reliable. Moreover the scattering cross section is small at large energies, making the modulation difficult to measure.

\begin{figure}[!h]
\begin{center}
\scalebox{0.8}{\includegraphics{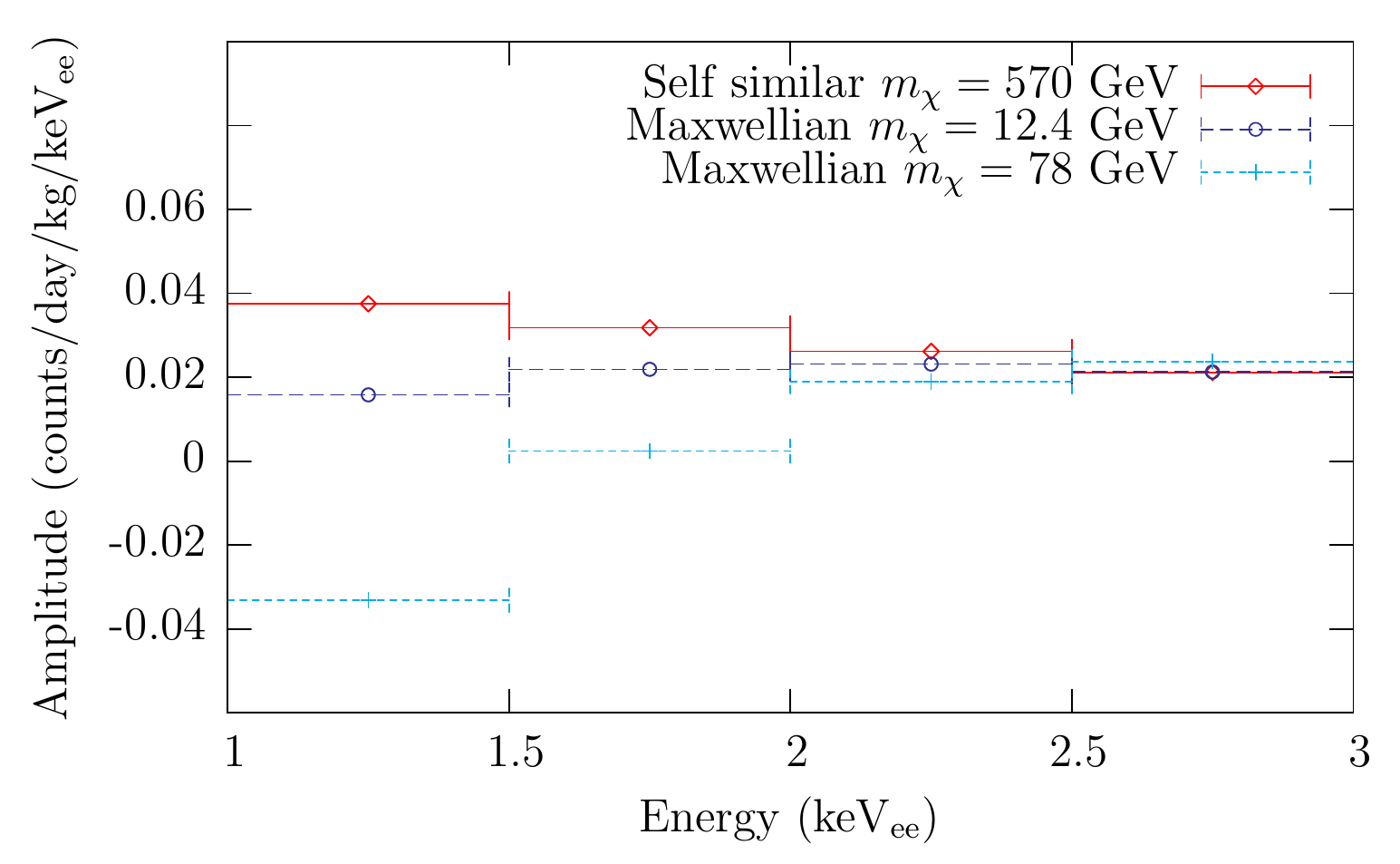}}
\end{center}
\caption{ The modulation amplitude at energies below the current DAMA threshold of 2 keV$_{\rm ee}$ (there is no data currently for $Q_{\rm det} < 2$ keV$_{\rm ee}$).  The self similar model (solid lines) shows an increase in amplitude with decrease in measured energy, while the Maxwellian models (broken lines) predict a smaller amplitude with decreasing energy owing to phase reversal at very low energies. The Maxwellian with $m_\chi$ = 78 GeV shows a negative amplitude in the lowest energy bin \label{fig8} }
\end{figure}

A phase reversal or non-sinusoidal behavior at low energies would be consistent with the Maxwellian, and inconsistent with the self similar model. As $F(Q) \sim 1$ at low energies, one may hope to use future data in these energy bins. The DAMA collaboration plans to add new photomultiplier tubes in order to achieve a lower energy threshold than the current threshold of 2 keV$_{\rm ee}$ \cite{dama1}. Let us suppose that future DAMA data includes measurements in the $1-1.5$ keV$_{\rm ee}$ and the $1.5-2$ keV$_{\rm ee}$ energy bins that are currently non-existent. Fig. \ref{fig8} shows the modulation amplitudes in these energy bins. The Maxwellian model with $m_\chi = 78$ GeV is the only one that shows a negative amplitude (i.e. the recoil rate in June is less than the recoil rate in December at these energies). The Maxwellian model with $m_\chi = 12.4$ GeV shows a small positive amplitude, while the self similar model with $m_\chi = 570$ GeV shows the largest positive modulation amplitude. These models can thus be distinguished provided sufficiently small error bars are achieved.

\subsubsection{Comparing the results of two different experiments.}

\cite{drees1, drees2} have described a technique of calculating the WIMP mass using the results of experiments with two different target nuclei, by comparing the moments of the distribution function. The great advantage of this approach is that it does not assume a form for the distribution function, and can thus be called model independent. Once the WIMP mass is determined in this way, it is possible to place constraints on the distribution function. Similarly, a lower bound on the WIMP mass from accelerator experiments can also constrain the form of the velocity distribution in the solar neighborhood. The parameter space near $m_\chi$ = 80 GeV is extremely sensitive to the presence of streams. We have checked that when the streams of the self similar model contribute as little as 5\% to the total dark matter density in the solar neighborhood (with the dominant stream contributing 2.35\% of the total), the expected recoil rate disagrees with the DAMA result in the 2-2.5 keV$_{\rm ee}$ bin at $> 6 \sigma$, for an assumed WIMP mass $m_\chi = 78$ GeV. This is so because in this energy bin (and for this mass), the minimum velocity required to produce a recoil is comparable to the velocity of the dominant stream and as a result, the stream is visible only during certain months of the year. As a result, the modulation due to the stream rivals that of the entire halo at this energy (see for example, \cite{savage} for a discussion of this effect). Thus provided the WIMP mass is determined to be $\sim 80$ GeV, the self similar model may be ruled out at high significance. Note also that only an experiment that is sensitive to the annual modulation will see this effect. The parameter space near $\sim 10$ GeV is far less sensitive to the presence of streams. This is because the minimum velocity required to produce recoils in the DAMA energy bins is often so large that all but the highest velocity streams are invisible. We find that stream fractions $\gtrsim$ 50\% are admissible when $m_\chi \sim 10$ GeV (see Fig. \ref{fig6}).

\subsubsection{Directional sensitivity.}

For the Maxwellian halo, the WIMP particles do not have a definite direction, instead they have a large velocity dispersion. As a result, the dark matter particles come predominantly from the direction of the sun's motion \cite{gelmini}. This is not the case for the self similar infall halo. Since the flows of the self similar model are cold (i.e. non-thermal), the WIMP flux depends on the velocity vectors of the flows, and the most intense flows are in a direction nearly opposite to that of the sun's motion \cite{copi, gelmini}. Thus a large WIMP wind due to the dominant flow can be easily distinguished from the prediction of the Maxwellian halo provided directional information is available.

\subsubsection{Measuring the average recoil rate in addition to the amplitude.}

\begin{figure}[!h]
\begin{center}
\scalebox{0.8}{\includegraphics{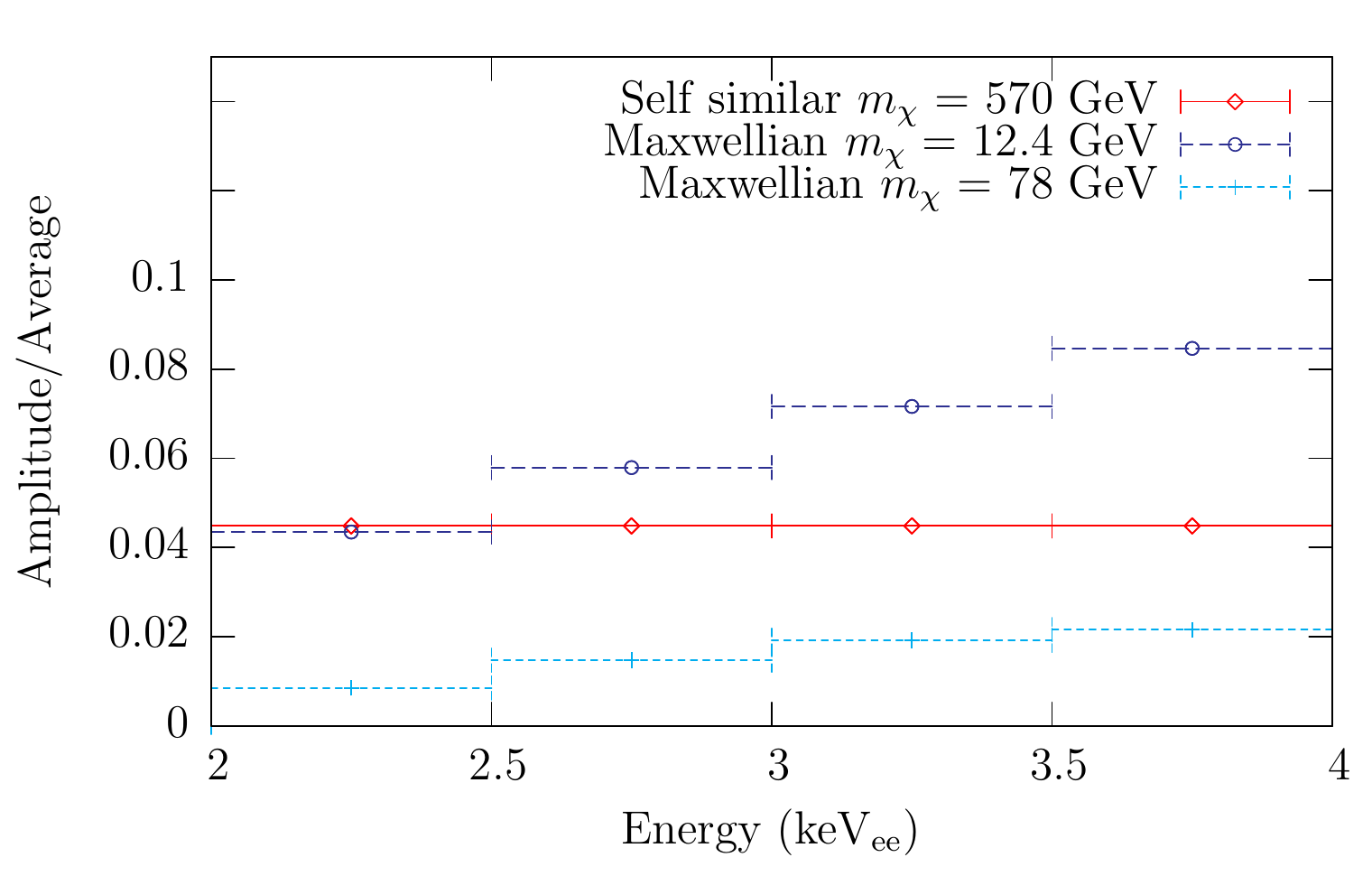}}
\end{center}
\caption{ Percentage modulation at different energies. The self similar model predicts a step like spectrum resembling a flat line at low energies. The Maxwellian models show a continuous variation with energy. \label{fig9} }
\end{figure}

As mentioned previously, the DAMA experiment does not measure the time averaged value of the recoil rate due to the large unidentified background. If this average can be measured, we have additional and complimentary information regarding the halo model. Fig. \ref{fig9} shows the percentage modulation which is the ratio of the modulation amplitude to the average value of the recoil rate, for the three models. This ratio has the advantage that its energy dependence is entirely due to the halo model. The self similar model which has discrete streams produces a flat spectrum at low energies, since all streams are visible. The Maxwellian models on the other hand show an increase in the percentage modulation with energy.

A detailed study of these ideas is left to future work. A major problem not considered in the present work is the incompatibility of the DAMA result with the null result of other experiments.  As seen from Figs. \ref{fig5} and \ref{fig6}, the self similar model does not help in bringing the DAMA result in agreement with other experiments. Inelastic scattering \cite{inel1, inel2, inel3, inel4, inel5} has been suggested as a possible solution to this discrepancy. In the inelastic scattering scenario, the internal energy of the WIMP is altered, thus only a small fraction of the kinetic energy of the incoming particle is transferred to the target. Inelastic scattering prefers heavy targets and for a suitable mass splitting (i.e. energy difference between the lowest and excited states of the WIMP), one can observe recoils with Iodine, and very few (or none) with a lighter element such as Germanium. In this scenario, the CDMS bounds are considerably relaxed, while the bounds from Xenon, CRESST, and KIMS are still relevant. Further study of this scenario, as well as an analysis of the excess events claimed by the CoGeNT experiment \cite{cogent} in low energy bins is left to future work. It is hoped that future dark matter detectors, particularly ones with directional sensitivity will be able to shed more light on the form of the local dark matter phase space distribution.

\acknowledgments{A.N. acknowledges financial support from the Bruce and Astrid McWilliams postdoctoral fellowship.  }


\begin{thebibliography}{99}

\bibitem{drukier}
A.K. Drukier, K. Freese, D.N. Spergel, Phys. Rev. D\textbf{33}, 3495 (1986)

\bibitem{freese}
K. Freese, J. Frieman, A. Gould, Phys. Rev. D\textbf{37}, 3388 (1988)

\bibitem{dama1}
R. Bernabei, for the DAMA collaboration, arXiv:1007.0595 (2010)

\bibitem{dama2}
R. Bernabei, for the DAMA collaboration, The European Physical Journal C\textbf{56}, 333 (2008)

\bibitem{cdms}
Z. Ahmed et al, for the CDMS collaboration, Phys. Rev. Lett. \textbf{102}, 011301 (2009)

\bibitem{xenon}
E. Aprile et al., for the Xenon100 collaboration, Phys. Rev. Lett., \textbf{105}, 131302 (2010)


\bibitem{cirelli}	
M. Cirelli, F. Iocco, P. Panci, JCAP10(2009)009

\bibitem{galli}
S. Galli, F. Iocco, G. Bertone, A. Melchiorri, Phys. Rev. D\textbf{80}, 023505 (2009)

\bibitem{arvi_reion}	
A. Natarajan, D.J. Schwarz, Phys. Rev. D\textbf{81}, 123510 (2010)

\bibitem{copi}
C.J. Copi, L.M. Krauss, Phys. Rev. D\textbf{63}, 043507 (2001) 

\bibitem{green}
A.M. Green, Phys. Rev. D\textbf{63}, 103003 (2001)

\bibitem{gelmini}
G. Gelmini, P. Gondolo, Phys. Rev. D\textbf{64}, 023504 (2001)

\bibitem{vergados}
J.D. Vergados, Phys. Rev. D\textbf{63}, 063511 (2001) 

\bibitem{ling}
F.S. Ling, P. Sikivie, S. Wick, Phys. Rev. D\textbf{70}, 123503 (2004)

\bibitem{form_factor1}
R.H. Helm, Phys. Rev., \textbf{104}, 1466 (1956)

\bibitem{form_factor2}
J.D. Lewin, P.F. Smith, Astropart. Phys., \textbf{6}, 87 (1996)

\bibitem{form_factor3}
G. Jungman, M. Kamionkowski, K. Griest, Phys. Rep. \textbf{267}, p.195 (1996)

\bibitem{chan}
R. Bernabei, for the DAMA collaboration, The European Physical Journal, C\textbf{53}, 205 (2008)

\bibitem{duffy}
L.D. Duffy, P. Sikivie, Phys. Rev. D\textbf{78}, 063508 (2008)

\bibitem{Fillmore}
J.A. Fillmore, P. Goldreich, Astrophys. J., \textbf{281}, 1 (1984)

\bibitem{bertschinger}
E. Bertschinger, Astrophysical Journal Supplement Series, \textbf{58}, 39 (1985)

\bibitem{sikivie1}
P. Sikivie, I.I. Tkachev, Y. Wang, Phys. Rev. Lett., \textbf{75},  2911 (1995) 

\bibitem{sikivie2}
P. Sikivie, I.I. Tkachev, Y. Wang, Phys. Rev. D\textbf{56}, 1863 (1997)

\bibitem{sikivie_ipser}
P. Sikivie, J.R. Ipser, Phys. Lett. B\textbf{291}, 288 (1992)

\bibitem{arvi1}
A. Natarajan, P. Sikivie, Phys. Rev. D\textbf{72}, 083513 (2005)

\bibitem{sikivie_crs}
P. Sikivie, Phys. Rev. D\textbf{60}, 063501 (1999)

\bibitem{arvi2}
A. Natarajan, P. Sikivie, Phys. Rev. D\textbf{73}, 023510 (2006)

\bibitem{kinney_sikivie}
W.H. Kinney, P. Sikivie, Phys. Rev. D\textbf{61}, 087305 (2000)

\bibitem{arvi3}
A. Natarajan, Advances in Astronomy, \textbf{2011}, id. 285346 (2011)



\bibitem{savage}
C. Savage, K. Freese, P. Gondolo, Phys. Rev. D\textbf{74}, 043531 (2006)

\bibitem{streams1}
D. Stiff, L.M. Widrow, J. Frieman, Phys. Rev. D\textbf{64}, 083516 (2001)

\bibitem{streams2}
K. Freese, P. Gondolo, H.J. Newberg, M. Lewis, Phys. Rev. Lett., \textbf{92}, 111301 (2004)

\bibitem{streams3}
P. Gondolo, G. Gelmini, Phys. Rev. D\textbf{71}, 123520 (2005)

\bibitem{streams4}
K. Freese, P. Gondolo, H.J. Newberg, Phys. Rev. D\textbf{71}, 043516 (2005)

\bibitem{caustics_evidence}
P. Sikivie, Phys. Lett. B\textbf{567}, 1 (2003)


\bibitem{savage2}
C. Savage, G. Gelmini, P. Gondolo, K. Freese, JCAP04(2009)010

\bibitem{fairbairn}
M. Fairbairn,  T. Schwetz, JCAP01(2009)037

\bibitem{gondolo_chan}
N. Bozorgnia, G. Gelmini, P. Gondolo,  arXiv:1006.3110 (2010)

\bibitem{mass_phase}
M.J. Lewis, K. Freese, Phys. Rev. D\textbf{70}, 043501 (2004)

\bibitem{drees1}
M. Drees, C.L. Shan, JCAP06(2007)011

\bibitem{drees2}
M. Drees, C.L. Shan, JCAP06(2008)012

\bibitem{inel1}
D. Smith, N. Weiner, Phys. Rev. D\textbf{64}, 043502 (2001); 

\bibitem{inel2}
S. Chang, G.D. Kribs, D. Tucker-Smith, N Weiner, Phys. Rev. D\textbf{79}, 043513 (2009); 

\bibitem{inel3}
J. Kopp, T. Schwetz, J. Zupan, JCAP02(2010)014

\bibitem{inel4}
D.S.M. Alves, M. Lisanti, J.G. Wacker, Phys. Rev. D \textbf{82}, 031901(R) (2010)

\bibitem{inel5}
P.J. Fox, G.D. Kribs, T.M.P. Tait, arXiv:1011.1910 (2010)

\bibitem{cogent}
C.E. Aalseth et al. for the CoGeNT collaboration, arXiv:1002.4703 (2010)

\end{thebibliography}
\end{document}